\documentclass[prb,twocolumn]{revtex4-1}
\usepackage{graphicx}
\usepackage{float}

\usepackage{amssymb}
\usepackage{amsmath}
\usepackage{bbold}
\usepackage{dcolumn}
\usepackage{colortbl}

\renewcommand{\vec}[1]{{\mathbf #1}}

\definecolor{LightGray}{gray}{0.85}

\bibpunct{[}{]}{,}{n}{}{}

\begin{document}


\title{Anderson transition for elastic waves in three dimensions}


\author{S.E. Skipetrov}
\email{sergey.skipetrov@lpmmc.cnrs.fr}
\affiliation{Univ. Grenoble Alpes, CNRS, LPMMC, 38000 Grenoble, France}
\author{Y.M. Beltukov}
\email{ybeltukov@gmail.com}
\affiliation{Department of Solid State Physics, Ioffe Institute, 194021 St. Petersburg, Russia}

\date{\today}

\begin{abstract}
We use two different fully vectorial microscopic models featuring nonresonant and resonant scattering, respectively, to demonstrate the Anderson localization transition for elastic waves in three-dimensional (3D) disordered solids. Critical parameters of the transition determined by finite-time and finite-size scaling analyses suggest that the transition belongs to the 3D orthogonal universality class. Similarities and differences between the elastic-wave and light scattering in strongly disordered media are discussed.
\end{abstract}

\maketitle

\section{Introduction}
\label{sec:intro}

Appearance of localized eigenstates and suppression of quantum or wave transport in a strongly disordered medium is a widespread physical phenomenon discovered by Philip Anderson 60 years ago \cite{anderson58,abrahams10}. Of special interest is the localization by three-dimensional (3D) disorder that takes place only when the latter is strong enough \cite{abrahams79}. A critical energy (or frequency) known as a ``mobility edge'' separates the range of energies (or frequencies) for which the eigenstates are localized from the rest of the spectrum. The localization transition taking place upon crossing a mobility edge is a subject of intense theoretical studies \cite{evers08}. It has been observed in experiments with electrons in doped semiconductors \cite{rosenbaum80,paalanen82}, vibrations in elastic networks \cite{hu08,cobus16}, and cold atoms in random potentials \cite{chabe08,jendr12}. Despite important experimental efforts, however, Anderson localization of light in 3D has not been unambiguously demonstrated up to now \cite{vanderbeek12,sperling16,skip16njp}. In addition, recent calculations show that the vector nature of light may hamper Anderson localization in a model of point scatterers \cite{skip14,bellando14}, raising a number of important questions. First, the extent to which the conclusion obtained for point scatterers may apply to disordered media composed of scatterers of finite size, is unclear \cite{escalante17}. Second, the relevance of results obtained for light to other types of vector waves remains unexplored. In particular, elastic waves (i.e., vibrations) in solids are directly concerned and one may wonder whether the theoretical results of Refs.\ \onlinecite{skip14,bellando14} may put in question the experimental observations of Refs.\ \onlinecite{hu08,cobus16}. Once the localization of elastic waves in 3D is firmly established, it is important to determine the universality class of the corresponding Anderson transition and to calculate its critical exponent which experiments are now attempting to measure \cite{cobus16}.

Various models of elastic wave systems have been shown to exhibit Anderson localization in 3D but most of them \cite{sheng94,chaudhuri10,mohthus10,pinski12,pinski12a,amir13,beltukov17} relied on the scalar approximation and are therefore not suitable to discuss the role of the vector character of vibrations. Calculations including the latter are scarce \cite{ludlam03,sepehrinia08} and predict critical properties that are different from those of scalar waves \cite{sepehrinia08}, in contradiction with general but rather formal theoretical arguments that seem to indicate that the vector character of vibrations is unlikely to modify the universality class of the localization transition \cite{john83,photiadis17}. Another model that takes into account the vector character of vibrations, is the model of a simple fluid with short-range interactions described by a truncated Lennard-Jones potential \cite{huang09}. It yields the same critical exponent of the localization transition for instantaneous normal modes as scalar vibrational models for both real (stable) and imaginary (unstable) frequencies. On the other hand, the vector and scalar percolation problems (also known as rigidity and connectivity percolation, respectively) belong to different universality classes and exhibit different critical exponents \cite{jacobs96}.

To clarify the role of the vector character of elastic waves in the context of Anderson localization, in this paper we report results obtained for two different models describing strong scattering of elastic waves in 3D with a full account for their vector nature. In the first model, the scattering is nonresonant and the localization transition takes place upon increasing the frequency of vibrations. This model is a direct extension of the scalar model of Ref.\ \onlinecite{beltukov17} and is similar to those employed to study heat transport by phonons in disordered materials \cite{sheng94,chaudhuri10}. The second model describes resonant scattering inducing Anderson localization in a narrow frequency band separated from the rest of the spectrum by two mobility edges, similarly to the scalar model of Ref.\ \onlinecite{skip16}. This is typical for disordered samples used in recent experiments on wave localization in elastic networks \cite{hu08,cobus16}. The critical properties of the first model are deduced from a finite-time scaling analysis whereas the second model is analyzed using a finite-size scaling technique. We show that in both cases and within numerical errors, the critical exponents of localization transitions coincide with those found for scalar waves. This suggests that the Anderson transition of elastic waves in 3D belongs to the orthogonal universality class.

\section{A model with nonresonant scattering}
\label{sec:nonres}

\subsection{Derivation of the model}

A model describing nonresonant scattering of elastic waves can be constructed by considering a cubic lattice of $N = L^3$ identical unit point masses (or atoms) interacting harmonically. The Cartesian components $u_m^{\alpha}$ ($\alpha = x,y,z$) of the vector displacement $\vec{u}_m(t)$ of an atom $m$ from its equilibrium position obey \cite{beltukov13}
\begin{eqnarray}
\ddot{u}_m^\alpha(t) = - \sum_{n \beta} M_{mn}^{\alpha\beta}u_n^\beta(t),
\label{eq:newton}
\end{eqnarray}
where $M_{mn}^{\alpha\beta}$ are elements of a $3N \times 3N$ dynamical matrix ${\hat M}$. The latter is composed of $N$ $3 \times 3$ blocks ${\hat M}_{mn}$, describing the coupling between Cartesian components of the displacement of the atom $m$ with those of the displacement of the atom $n$. The dynamical matrix of a mechanically stable system should be positive semidefinite (i.e. its eigenvalues $\omega^2$ should be nonnegative) and hence it can be represented as ${\hat M} = {\hat A} {\hat A}^T$, that is
\begin{eqnarray}
M_{mn}^{\alpha\beta} = \sum_{k\gamma} A_{mk}^{\alpha\gamma}A_{nk}^{\beta\gamma}.
\label{eq:AAT}
\end{eqnarray}
Disorder is introduced in the model by assuming that the elements $A_{mn}^{\alpha\beta}$ of off-diagonal matrix blocks ($m \ne n$) describing interactions between nearest-neighbor atoms, are real independent, zero-mean Gaussian variables with variances {$\Omega^2$}. $A_{mn}^{\alpha\beta} = 0$ for atoms $m$ and $n$ that are not nearest neighbors. The diagonal blocks $A_{mm}^{\alpha\beta}$ are obtained by a sum rule $A_{mm}^{\alpha\beta} = -\sum_{n\ne m} A_{nm}^{\alpha\beta}$, which ensures that the total potential energy is invariant upon translation of the system as a whole.
{We will measure the frequency $\omega$ in units of $\Omega$ and the time $t$ in units of $1/\Omega$ from here on.
}


\begin{figure}
\includegraphics[width=0.99\columnwidth]{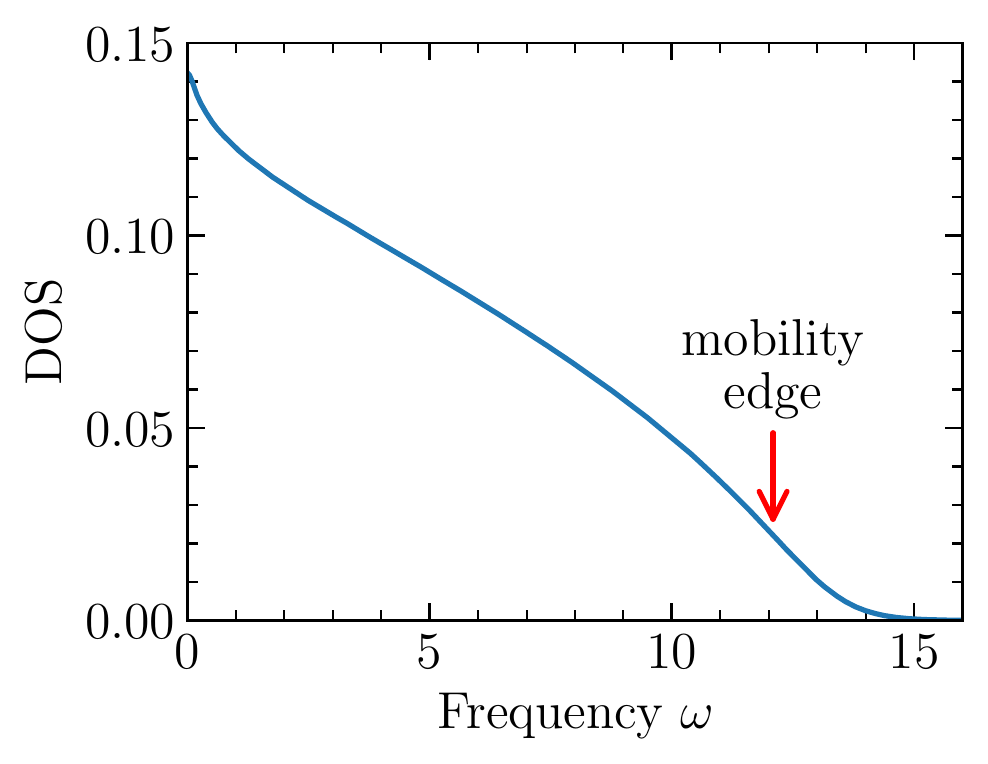}\\
\caption{\label{fig_dos} {Density of states of the model with nonresonant scattering. The red arrow indicates the position of the mobility edge determined in Sec.\ \ref{sec:fts}.}}
\end{figure}

{Figure \ref{fig_dos} shows the density of states (DOS) of the model defined by Eqs.\  (\ref{eq:newton}) and (\ref{eq:AAT}). It is simply a probability density of a random variable $\omega$, with $\omega^2$ being the eigenvalues of the random matrix ${\hat M}$. As we will show in Sec.\ \ref{sec:fts}, Anderson localization takes place for frequencies $\omega$ exceeding a critical frequency (mobility edge) $\omega_c \simeq 12$ shown  by a vertical arrow in Fig.\ \ref{fig_dos}. The DOS vanishes exponentially beyond some $\omega_{\mathrm{max}} \simeq 15$ determined by the statistics of the random matrix ${\hat A}$. In this respect, the model is quite particular because it assumes that the properties of the system are dominated by the disorder and neglects the ``regular'' part of ${\hat M}$ that would describe the system in the absence of disorder. As a consequence, the low-frequency behavior of the DOS in Fig.\ \ref{fig_dos} does not show the shape expected in a homogeneous medium $\mathrm{DOS}(\omega) \propto \omega^2$ (Debye law) for $\omega \to 0$. This is due to vanishing of the Young modulus and, consequently, vanishing of both the rigidity of the lattice and the zero-frequency sound velocity $v$ in our model \cite{beltukov13}. Then, the Ioffe-Regel criterion $k(\omega) \ell(\omega) \lesssim 1$, where $k(\omega)$ is the wavenumber and $\ell(\omega)$ is the scattering mean free path, is obeyed already for $\omega = 0$. The model (\ref{eq:AAT}) can be made more realistic by defining ${\hat M}$ as a sum of ${\hat A} {\hat A}^T$ and an additional matrix ${\hat M}_0$ corresponding to the lattice without disorder. This introduces a non-zero rigidity of the lattice \cite{beltukov13} and produces a Debye-like behavior of the low-frequency spectrum in Fig.\ \ref{fig_dos}: $\mathrm{DOS}(\omega) \propto \omega^2$, which extends from $\omega = 0$ to some finite $\omega \simeq \omega_{\mathrm{IR}}$. Here $\omega_{\mathrm{IR}}$ is the Ioffe-Regel frequency obeying $k(\omega_{\mathrm{IR}}) \ell(\omega_{\mathrm{IR}}) \simeq 1$. $\omega_{\mathrm{IR}}$ can be made arbitrary small by decreasing the magnitude of ${\hat M}_0$ and has nothing to do with the mobility edge $\omega_c$ that is hardly affected by the introduction of ${\hat M}_0$. Therefore, the Ioffe-Regel criterion $k(\omega) \ell(\omega) \simeq 1$ is not a good condition of Anderson localization in the model defined by Eqs.\  (\ref{eq:newton}) and (\ref{eq:AAT}).
A discrepancy between $\omega_{\mathrm{IR}}$ and $\omega_c$ for vibrational modes has been also noticed for other models of disordered solids \cite{taraskin99}.
}

{To study the localization transition,} we compute the spreading of an initial excitation of the left half of the system [$\dot{u}^\alpha_m(0)$ are taken to be Gaussian random numbers for $x_m<0$ and 0 elsewhere; all $u^\alpha_m(0) = 0$] by extending the numerical approach of Ref.\ \onlinecite{beltukov17} to the vector case. The spreading is quantified by a frequency-resolved penetration depth $X(\omega, t)$:
\begin{eqnarray}
X^2(\omega, t) = \frac{1}{\phi_0(\omega)}\int\limits_0^{\infty} x \phi(\omega, x,t) dx,
\label{depth}
\end{eqnarray}
where
$\phi(\omega, x,t) = \sum_m E_m(\omega, t) \delta(x - x_m)$
is the 1D energy density,
\begin{eqnarray}
E_m(\omega, t) &=&  \frac{1}{2} \sum_\alpha \dot{u}_m^\alpha(\omega, t)^2
\nonumber \\
&+&
\frac{1}{2}\sum_{n\alpha\beta}M_{mn}^{\alpha\beta}u_m^\alpha(\omega, t)u_n^\beta(\omega, t)
\end{eqnarray}
is the energy of a quasimonochromatic excitation localized on the atom $m$, and $\phi_0(\omega)=2E(\omega)/L$ is the average initial energy density in the left half of the system. The windowed Fourier transform of $u_m^{\alpha}(t)$ is defined as
\begin{eqnarray}
u_m^{\alpha}(\omega, t) = 2\int\limits_{-\tau}^{\tau} u_m^{\alpha}(t-t') W(t')\cos(\omega t')dt'
\end{eqnarray}
with a window function $W(t) = (2\pi\tau)^{-1/2}\cos(\pi t/2\tau)$. We use $L = 200$ and average $X^2(\omega,t )$ over 10 realizations of disorder.
{This relatively small number of realizations turns out to be sufficient to suppress the statistical fluctuations of $X^2(\omega,t )$ thanks to the additional implicit averaging over the transverse profile of the spreading excitation in the $(y, z)$ plane and over the many atoms that are initially excited in the left half of the system. Indeed, for a 3D system of $L \times L \times L$ point masses, even a single realization of disorder yields $X^2(\omega,t )$ that is effectively averaged over $L^2$ different linear chains of effective length $\sqrt{Dt}$ in diffuse regime or $\xi$ in the regime of Anderson localization. Here $D$ and $\xi$ are the diffusion coefficient and the localization length, respectiveley. The results would be self-averaging in the limit $L  \to \infty$, $t \to \infty$ but additional configurational averaging is necessary for finite $L$ and $t$.
}

\begin{figure}
\includegraphics[width=0.99\columnwidth]{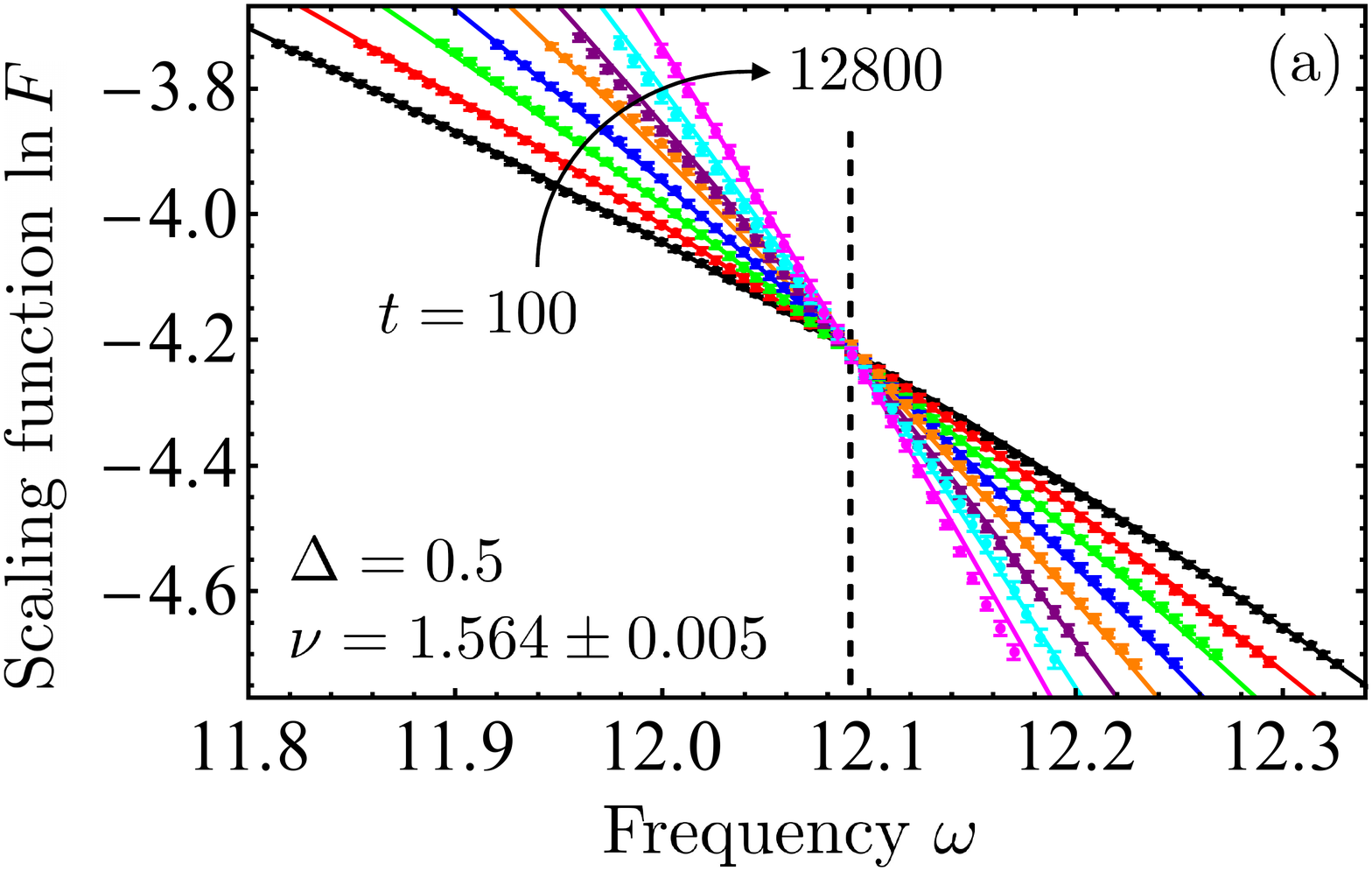}\\
\vspace{-7mm}
\includegraphics[width=0.99\columnwidth]{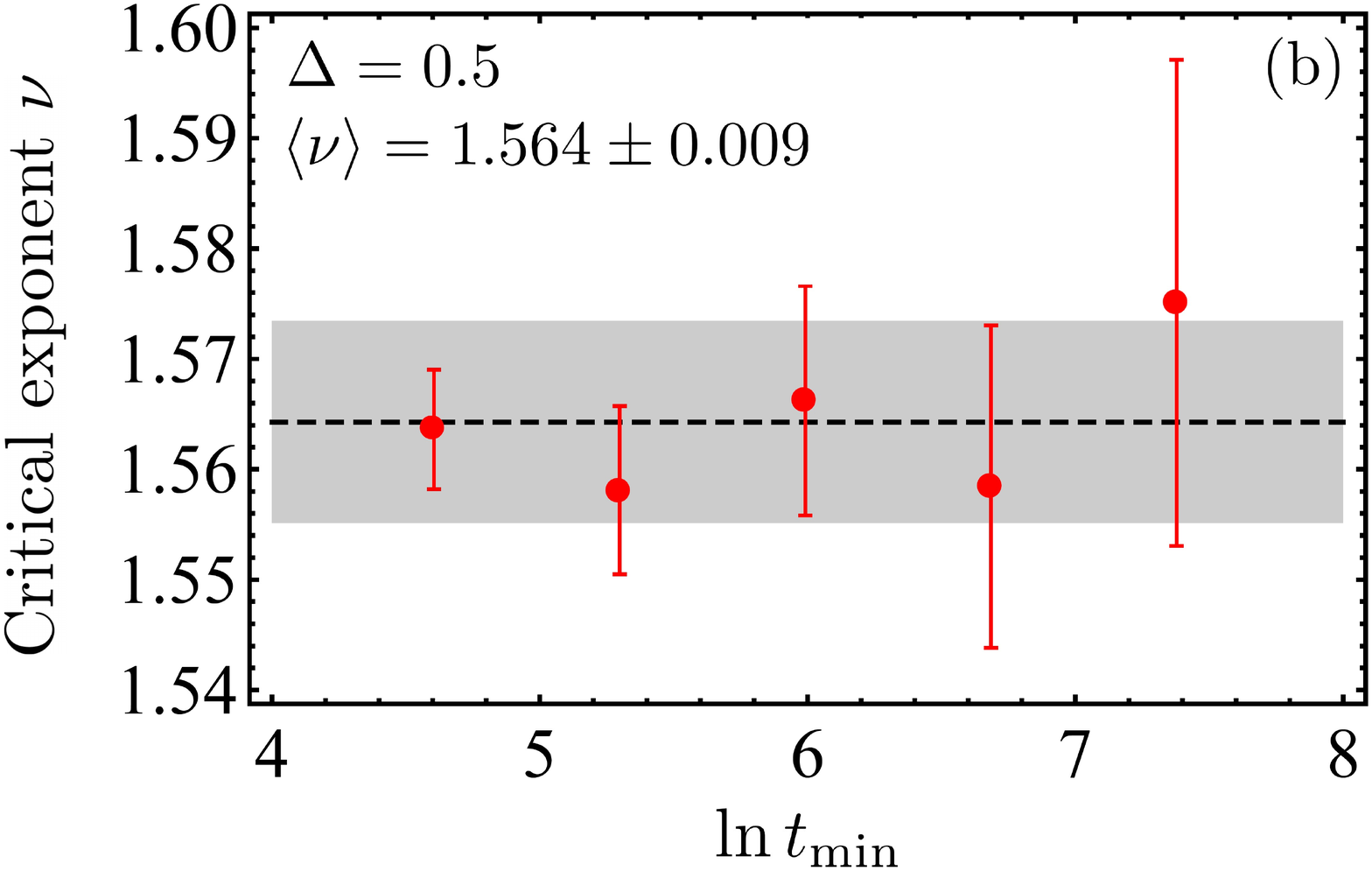}
\vspace{-5mm}
\caption{\label{fig_fit_2100} Results for the model with nonresonant scattering. (a) An example of a fit with $m_1 = 2$, $n_1 = 1$, and $m_2 = n_2 = 0$. Symbols with error bars show numerical data for different times $t = 100$ (black), 200 (red), 400 (green), 800 (blue), 1600 (orange), 3200 (purple), 6400 (cyan), 12800 (magenta). Solid lines are polynomial fits, a dashed vertical line shows the mobility edge $\omega_c \simeq 12.09$. (b) The critical exponent $\nu$ extracted from the fits similar to the one in panel (a) using only the data corresponding to $t \geq t_{\mathrm{min}}$, as a function of $t_{\mathrm{min}}$. Error bars are equal to one standard deviation of $\nu$. The dashed horizontal line shows the average value of $\nu$, the grey area shows the error of the average.}
\end{figure}

\subsection{Finite-time scaling of the penetration depth}
\label{sec:fts}

Similarly to the scalar case \cite{beltukov17}, $\langle X^2(\omega, t) \rangle$ grows linearly with time for $\omega < \omega_c \simeq 12$ but acquires a tendency to saturation at higher frequencies. This identifies $\omega_c$ as a mobility edge, which turns out to be shifted to a higher frequency compared to the scalar model. The critical behavior in the vicinity of $\omega_c$ can be studied by analyzing a scaling function $F(\omega, t ) = \rho(\omega)^{2/3} \langle X^2(\omega, t) \rangle/t^{2/3}$, where $\rho(\omega)$ is the density of vibrational states \cite{beltukov17}. Under the hypothesis of one-parameter scaling, $F$ is assumed to depend on a single relevant variable $\psi_1 = [t/\rho(\omega)]^{1/3\nu} f_1(w)$ and possibly on an additional irrelevant variable $\psi_2 = [t/\rho(\omega)]^{y/3} f_2(w)$, where $w = (\omega-\omega_c)/\omega_c$, $\nu$ is the critical exponent of the localization transition, and $y < 0$. $\ln F$ is expanded in power series as a function of $\psi_j$ up to orders $n_j$, whereas the auxiliary functions $f_j$ are expanded in powers of $w$ up to orders $m_j$. These expansions are used to fit the numerical data obtained in the vicinity of $\omega_c$ for different frequencies $\omega$ and different times $t$. The range of data used for the fits is restricted to $\ln F$ within a range $\ln \tilde{F}_c \pm \Delta$, with $\ln \tilde{F}_c$ being the approximate crossing point of dependencies $\ln F(\omega)$ obtained for different $t$. The quality of our numerical data allows us to use $\Delta$ as small as $\Delta = 0.5$ for which fits of acceptable quality are obtained with $m_1 = 2$, $n_1 = 1$, and $m_2 = n_2 = 0$, see Fig.\ \ref{fig_fit_2100}(a) \cite{note1}.

To test the stability of the fit presented in Fig.\ \ref{fig_fit_2100}(a) and obtain a better estimate of the accuracy of the value of the critical exponent following from it, we repeat the fit for data corresponding to long times $t \geq t_{\mathrm{min}}$ only.
{$t_{\mathrm{min}}$ is a free parameter that we vary from the shortest time $t = 100$ used in our calculations to $t = 1600$, which is still considerably shorter than the longest time $t = 12800$ in Fig.\ \ref{fig_fit_2100}(a). The best-fit $\omega_c$ and $\nu$ should not depend on $t_{\mathrm{min}}$, whereas their statistical errors are likely to grow when $t_{\mathrm{min}}$ increases because a longer $t_{\mathrm{min}}$ implies using a smaller fraction of available numerical data (i.e., only the data corresponding to $t \geq t_{\mathrm{min}}$). The independence of $\nu$ from $t_{\mathrm{min}}$ is confirmed by Fig.\ \ref{fig_fit_2100}(b) where we show}
the best-fit critical exponent $\nu$ as a function of $t_{\mathrm{min}}$. Averaging over $t_{\mathrm{min}}$ yields $\langle \nu \rangle = 1.564 \pm 0.009$, which coincides, within error bars, with $\langle \nu \rangle = 1.57 \pm 0.02$ found in the scalar version of the same model \cite{beltukov17}.

\section{A model with resonant scattering}
\label{sec:res}

The disordered elastic media in which Anderson localization of vibrations has been observed in experiments \cite{hu08,cobus16}, were composed of individual scattering units having strong scattering resonances (identical aluminum beads brazed together). The applicability of the nonresonant scattering model considered above to such a medium is questionable and one may wonder whether localization transitions in these two systems belong to the same universality class. To answer this question, we consider a model in which scattering is due to $N$ identical point-like resonant scatterers (resonance frequency $\omega_0$, resonance width $\Gamma_0$)
{embedded in the infinite space filled with a homogeneous and isotropic elastic medium.}

\subsection{Derivation of the model}

{Assume that the scatterers are} randomly distributed in a spherical region of space
{of radius $R$ and volume $V$} with an average {number} density $\rho = N/V$.
{Propagation of an elastic wave in the homogeneous space between the scatterers is described by the elastic wave equation \cite{landau}
\begin{eqnarray}
\rho_0 \omega^2 u^{\alpha}(\vec{r}, \omega) &+& \sum\limits_{\beta, \gamma, \zeta}
\frac{\partial}{\partial r^{\beta}} \left[ c_{\alpha \beta \gamma \zeta} \frac{\partial}{\partial r^{\gamma}} u^{\zeta}(\vec{r}, \omega) \right] \nonumber \\
&=& -f^{\alpha}(\vec{r}, \omega),
\label{waveeq}
\end{eqnarray}
where $\vec{u}(\vec{r}, \omega)$ is the Fourier transform of the displacement field $\vec{u}(\vec{r}, t)$, $\rho_0$ is the mass density of the medium, $c_{\alpha \beta \gamma \zeta} = \lambda_0
\delta_{\alpha \beta} \delta_{\gamma \zeta}+
\mu_0 (\delta_{\alpha \gamma} \delta_{\beta \zeta} + \delta_{\alpha \zeta} \delta_{\beta \gamma})$
is its elasticity tensor, $\lambda_0$ and $\mu_0$ are Lam\'{e} parameters, $\vec{f}(\vec{r}, \omega)$ is the Fourier transform of the force field $\vec{f}(\vec{r}, t)$, and Greek superscripts run over the Cartesian components of the corresponding vectors: $\alpha$, $\beta$, $\gamma$, $\zeta = x, y, z$.}

{Solutions of Eq.\ (\ref{waveeq}) can be classified as compressive (longitudinal) and shear (transverse) waves that travel with velocities $c_p = [(\lambda_0 + 2 \mu_0)/\rho_0]^{1/2}$ and $c_s = (\mu_0/\rho_0)^{1/2}$, respectively \cite{landau}. A $3 \times 3$ elastic Green's tensor ${\hat{\cal G}}$ describes the response of the medium to a point-like excitation. Its component ${\cal G}^{\alpha \eta} (\vec{r}, \vec{r}', \omega)$ gives the (Fourier component of) displacement at location $\vec{r}$ in direction $\alpha$ due to a unit force $\vec{f}$ acting in direction $\eta$ at a point $\vec{r}'$. It obeys an equation following directly from Eq.\ (\ref{waveeq}):
\begin{eqnarray}
\rho_0 \omega^2 {\cal G}^{\alpha \eta}(\vec{r},\vec{r}',\omega) &+& \sum\limits_{\beta, \gamma, \zeta}
\frac{\partial}{\partial r^{\beta}} \left[ c_{\alpha \beta \gamma \zeta} \frac{\partial}{\partial r^{\gamma}} {\cal G}^{\zeta \eta}(\vec{r},\vec{r}',\omega) \right]
\nonumber \\
&=& -\delta_{\alpha \eta} \delta(\vec{r}-\vec{r}').
\label{greeneq}
\end{eqnarray}
The solution of this equation can be written as \cite{ben81,snieder02}
\begin{eqnarray}
{\hat{\cal G}}(\vec{r}, \vec{r}',\omega) &=& \frac{\omega}{12 \pi \rho_0 c_p^3} \left\{
\frac{e^{i k_p \Delta r}}{k_p \Delta r} \left[ \mathbb{1}
- {\hat F}(k_p \Delta \vec{r}) \right] \right.
\nonumber \\
&+& \left. 2 \left( \frac{c_p}{c_s} \right)^3
\frac{e^{i k_s \Delta r}}{k_s \Delta r} \left[ \mathbb{1}
+ \frac12 {\hat F}(k_s \Delta \vec{r}) \right] \right\},
\label{green0}
\end{eqnarray}
where $k_{p,s} = \omega/c_{p,s}$, $\Delta \vec{r} = \vec{r}-\vec{r}'$, and we defined an auxiliary tensor function ${\hat F}(\vec{v}) = [\mathbb{1} - 3 (\vec{v} \otimes \vec{v})/v^2] (1 + 3i/v - 3/v^2)$.}

{To describe the multiple scattering of elastic waves by an ensemble of $N$ identical resonant point scatterers at positions $\{ \vec{r}_m \}$ ($m = 1, \ldots, N$), we generalize the method developed by Foldy \cite{foldy45} and Lax \cite{lax51} for scalar waves. The Cartesian components $u_m^{\alpha}$ ($\alpha = x$, $y$, $z$) of displacements $\vec{u}_m$ of the scatterers obey
\begin{eqnarray}
u_m^{\alpha}(\omega) = v_m^{\alpha}(\omega) + \alpha_0(\omega) \sum\limits_{n \ne m, \beta}
{\cal G}^{\alpha \beta}(\vec{r}_m, \vec{r}_n,\omega) u_n^{\beta}(\omega),\;\;\;\;\;
\label{foldylax}
\end{eqnarray}
where $v_m^{\alpha}(\omega)$ is wave field at $\vec{r}_m$ due to the force field $\vec{f}(\vec{r},\omega)$  in the absence of scatterers (or the ``incident wave'' in the language of the scattering theory) and the second term on the right-hand side is a sum of wave fields scattered by all other scatterers. The function $\alpha_0(\omega)$ (a ``polarizability'' in the optical language) describes the response of individual scatterers. For an isotropic scatterer with resonance frequency $\omega_0$ and resonance width $\Gamma_0 \ll \omega_0$, it can be written as
\begin{eqnarray}
\alpha_0(\omega) = B \frac{(-\Gamma_0/2)}{\omega - \omega_0 + i \Gamma_0/2}
\label{polar}
\end{eqnarray}
with a prefactor $B = 12 \pi \rho_0 c_p^3/\omega [1 + 2 (c_p/c_s)^3]$ following from the optical theorem.
Equation (\ref{foldylax}) holds only for infinitesimal displacements $\vec{u}_m$ that do not modify the positions $\vec{r}_m$ of scatterers.
}

{
By grouping all $u_m^{\alpha}$ in a single vector $| u \rangle = (u_1^{x},u_1^{y},u_1^{z}, \ldots, u_N^{x},u_N^{y},u_N^{z})^{T}$ (and similarly for $v_m^{\alpha}$), we rewrite Eq.\ (\ref{foldylax}) in a vector form:
\begin{eqnarray}
| u(\omega) \rangle = | v(\omega) \rangle + \frac{\alpha_0(\omega)}{B} \left[ {\hat G}(\omega) - i \mathbb{1} \right] | u(\omega) \rangle,
\label{foldylaxvector}
\end{eqnarray}
where we use the Dirac bra-ket notation. The $3N \times 3N$ ``Green's matrix'' ${\hat G}$ it is composed of $N \times N$ blocks ${\hat G}_{mn}$ of size $3 \times 3$ each, describing coupling between Cartesian components of displacements of scatterers $m$ and $n$:}
\begin{eqnarray}
{\hat G}_{mn}(\omega) &=&
{i \delta_{mn} \mathbb{1} + (1 - \delta_{mn}) B {\cal G}(\vec{r}_m, \vec{r}_n, \omega)
}
\nonumber \\
&=& i \delta_{mn} \mathbb{1} + \frac{1 - \delta_{mn}}{1 + 2 (c_p/c_s)^3}
\nonumber \\
&\times& \left\{
\vphantom{\left( \frac{c_p}{c_s} \right)^3} \frac{e^{i k_p r_{mn}}}{k_p r_{mn}} \left[ \mathbb{1}
- {\hat F}(k_p \vec{r}_{mn}) \right] \right.
\nonumber \\
&+& \left. 2 \left( \frac{c_p}{c_s} \right)^3
\frac{e^{i k_s r_{mn}}}{k_s r_{mn}} \left[ \mathbb{1}
+ \frac12 {\hat F}(k_s \vec{r}_{mn}) \right] \right\},\;\;\;\;
\label{green}
\end{eqnarray}
where {$\vec{r}_{mn} = \vec{r}_m - \vec{r}_n$.}
The Green's matrix ${\hat G}$ is the central object describing all the peculiarities of wave scattering by an ensemble of resonant point scatterers. Equation (\ref{green}) gives its form for elastic waves; similar expressions for scalar and electromagnetic waves have been derived and analyzed previously \cite{skip14,skip15,skip16}. We notice that Eq.\ (\ref{green}) depends on very few parameters: the frequency $\omega$ and the speeds of compressive and shear waves $c_p$ and $c_s$, respectively.

In general, Eq.\ (\ref{foldylaxvector}) is difficult to analyze, but we can simplify it by making use of the strongly resonant nature of scattering. Indeed, the narrow resonance condition $\Gamma_0 \ll \omega_0$ implies that all the interesting phenomena due to strong scattering can take place only for $\omega \simeq \omega_0$, for which the scattering by individual scatterers is strong. This allows us to replace ${\hat G}(\omega)$ in Eq.\ (\ref{foldylaxvector}) by ${\hat G}(\omega_0)$. Such an approximation is equivalent to neglecting the times $R/c_{p,s}$  that compressive and share waves need to propagate across the disordered system and will allow us to describe slow processes taking place on large time scales $\tau \gg R/c_{p,s}$, which is the case for the slow decay of localized quasi-modes that we are intended to study.

{The matrix ${\hat G}(\omega_0)$ is a non-Hermitian matrix with complex eigenvalues $\Lambda_n$ and right and left eigenvectors $| R_n \rangle$ and $\langle L_n|$ obeying
\begin{eqnarray}
&&{\hat G}(\omega_0) | R_n \rangle = \Lambda_n | R_n \rangle, \label{rightev}\\
&&\langle L_n | {\hat G}(\omega_0) = \langle L_n | \Lambda_n, \label{leftev}\\
&&\langle L_m | R_n \rangle = \delta_{mn}. \label{ortho}
\label{lreigen}
\end{eqnarray}
$| R_n \rangle$ and $\langle L_n|$ form a biorthogonal basis in which any solution $| u \rangle$ of Eq.\ (\ref{foldylaxvector}) with ${\hat G}(\omega)$ replaced by ${\hat G}(\omega_0)$ can be represented as
\begin{eqnarray}
| u(\omega) \rangle = \sum\limits_n A_n(\omega) | R_n \rangle,
\label{represent}
\end{eqnarray}
where the coefficients $A_n$ are found by substituting the expansion (\ref{represent}) in Eq.\ (\ref{foldylaxvector}), applying Eq.\ (\ref{rightev}), multiplying both sides of the resulting equation by $\langle L_m |$ from the left, and applying Eq.\ (\ref{ortho}):
\begin{eqnarray}
A_n(\omega) = \frac{(-\Gamma_0/2)B}{\alpha_0(\omega)}\, \frac{\langle L_n | v(\omega) \rangle}{\omega - \omega_n + i \Gamma_n/2}
\label{coeff}
\end{eqnarray}
with $\omega_n = \omega_0 - (\Gamma_0/2) \mathrm{Re} \Lambda_n$ and $\Gamma_n = \Gamma_0 \mathrm{Im} \Lambda_n$.
}

{The physical meaning of the eigenvectors and eigenvalues of the matrix ${\hat G}(\omega_0)$ now becomes clear. For a short-pulse excitation $| v(t) \rangle \propto \delta(t)$, the Fourier transform of $| u(\omega) \rangle$ is
\begin{eqnarray}
| u(t) \rangle = \sum\limits_n A_n(t) | R_n \rangle
\label{representtime}
\end{eqnarray}
with $A_n(t) \propto \exp(-i \omega_n t - \Gamma_n t/2)$.
Thus, the eigenvectors $| R_n \rangle$ correspond to ``quasi-modes'' of our wave system, where the prefix ``quasi-'' refers to the fact that, in contrast to the modes of a closed system described by a Hermitian Hamiltonian, the quasi-modes decay in time with decay rates $\Gamma_n$. The latter are due to the openness of our wave system: the waves can freely escape from the region of space $V$ occupied by the scatterers, causing the leakage of wave energy and decay of any excitation that was initially created inside $V$. The concept of quasi-modes is particularly useful and physically meaningful when $\Gamma_n \ll \omega_n$, which will be the case in the following. Because the eigenvectors and the eigenvalues of the matrix ${\hat G}(\omega_0)$ correspond to the quasi-modes and their complex frequencies, ${\hat G}(\omega_0)$ plays the role of an effective non-Hermitian Hamiltonian for elastic waves in the considered system of resonant point scatterers.
}

\subsection{Finite-size scaling of the distribution of Thouless conductance}
\label{secfss}

{Similarly to what has been done for scalar waves \cite{skip16} and light \cite{skip18}, the localization transition for elastic waves in an ensemble of resonant point scatterers can be studied by analyzing the eigenvalues $\Lambda_n$ of the Green's matrix ${\hat G}(\omega_0)$. This may appear counterintuitive because the spatial localization is a property of eigenvectors $| R_n \rangle$ and not of eigenvalues, so that it would be logical to analyze the former instead of the latter. There is no doubt that the analysis of the spatial structure of eigenvectors $| R_n \rangle$ is a more direct way to study Anderson localization. However, one should take into account that such an analysis is much more involved both analytically, because very little is known about the properties of eigenvectors of non-Hermitian matrices, and numerically, because finding the eigenvectors of a large matrix requires considerably more computational resources than finding only the eigenvalues. The last limitation is crucial for us because an accurate characterization of a localization transition, including finding the critical frequency and, especially, the critical exponent, requires averaging over many random configurations of scatterers $\{ \vec{r}_m \}$ for large numbers of scatterers $N \gg 1$, and turns out to be a quite demanding computational task even when only the eigenvalues of ${\hat G}(\omega_0)$ are computed. Therefore, in the present work we restrict our analysis to the eigenvalues of ${\hat G}(\omega_0)$ and leave the statistics of eigenvectors for a future study. Nonetheless, we will discuss some properties of eigenvectors in Sec.\ \ref{secipr}, though without attempting a statistical analysis.
}

\begin{figure}
\includegraphics[width=0.99\columnwidth]{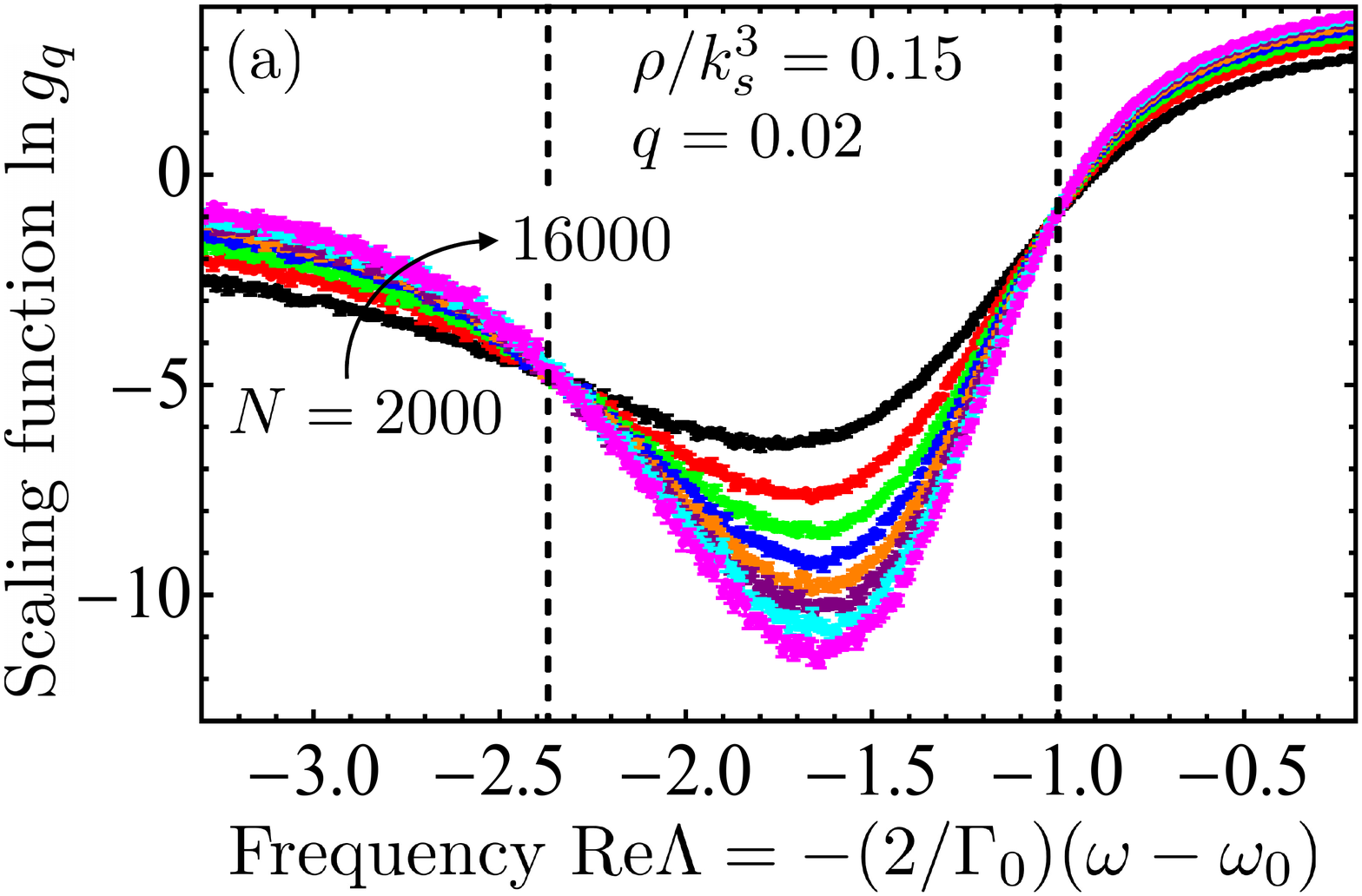}\\
\vspace{-7mm}
\includegraphics[width=0.99\columnwidth]{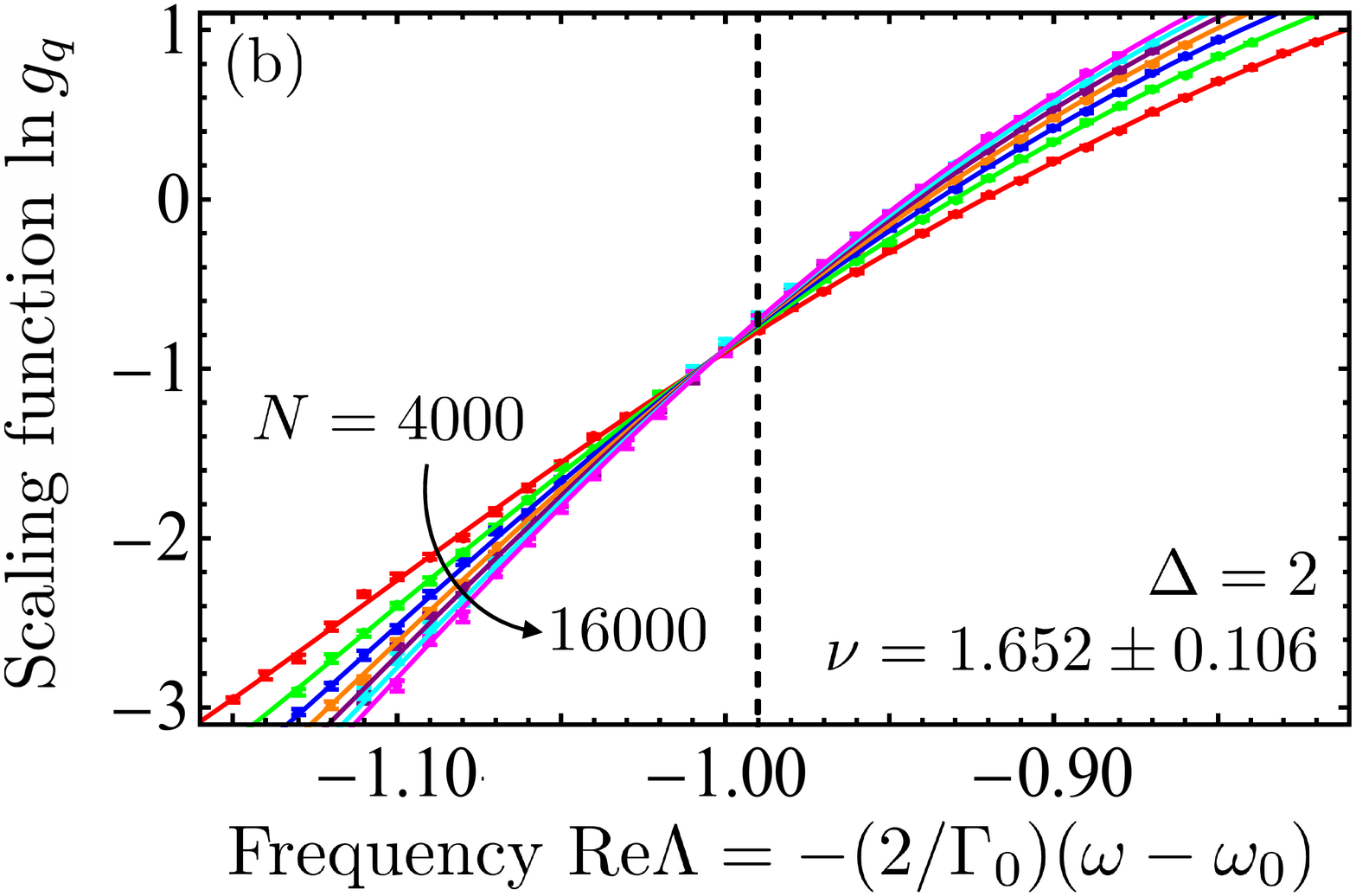}
\vspace{-5mm}
\caption{\label{fig_percentiles} Analysis of the model with resonant scattering. (a) Second percentile of the distribution $p(\ln g)$ at a fixed density $\rho$ of scatterers in a sphere for 8 different total numbers of scatterer $N = 2000$ (black), 4000 (red), 6000 (green), 8000 (blue), 10000 (orange), 12000 (purple), 14000 (cyan), 16000 (magenta). Vertical dashed lines show approximate positions of mobility edges where lines corresponding to different $N$ all cross. (b) A polynomial fit to the data of panel (a) in the vicinity of one of the mobility edges ($m_1 = n_1 = 3$, $m_2 = n_2 = 1$). The best-fit position of the mobility edge $\mathrm{Re} \Lambda_c \simeq -0.99$ is shown by the dashed vertical line; the critical exponent $\nu$ following from the fit is printed on the graph.}
\end{figure}

\begin{figure}
\includegraphics[width=0.99\columnwidth]{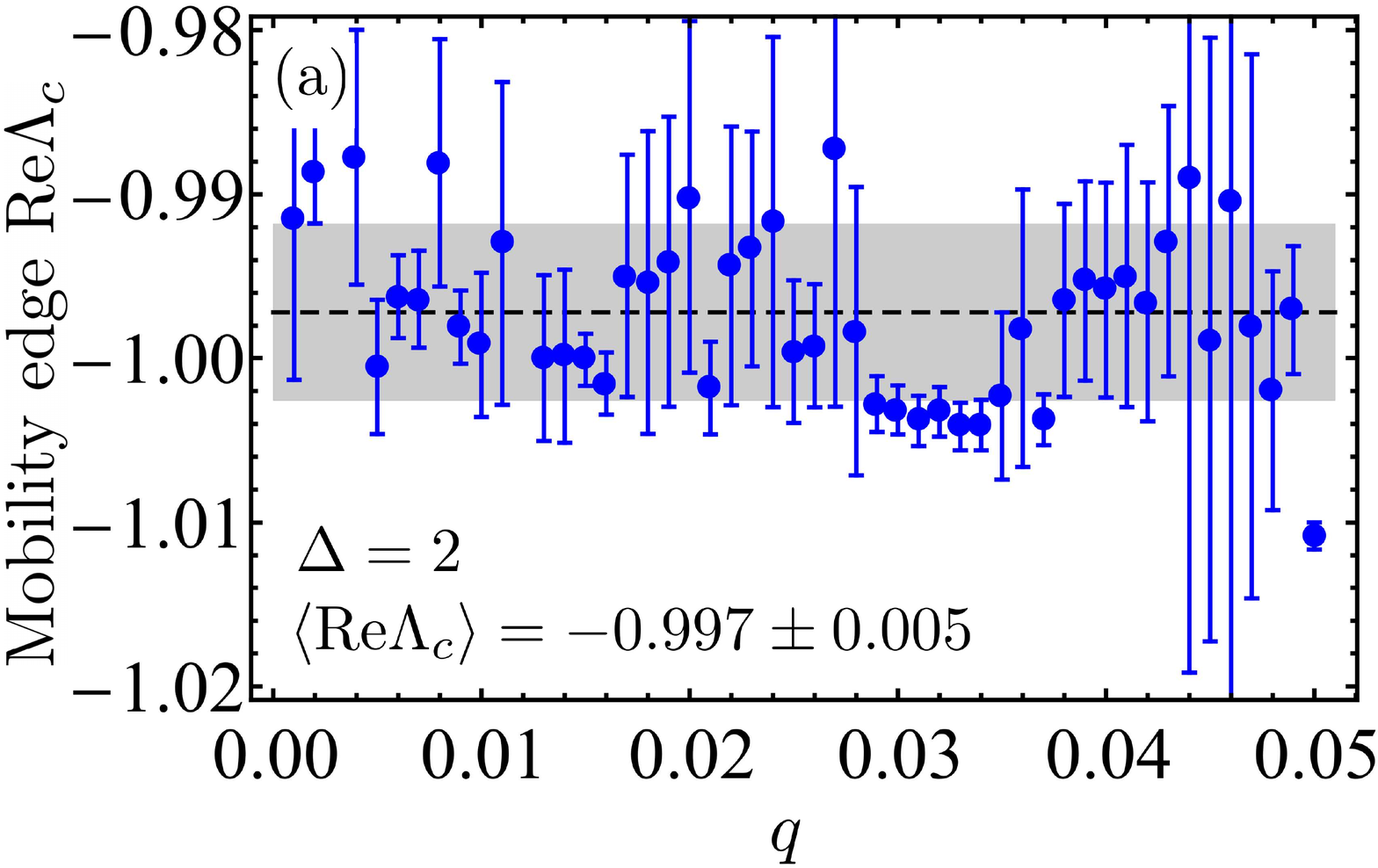}\\
\vspace{-7mm}
\includegraphics[width=0.99\columnwidth]{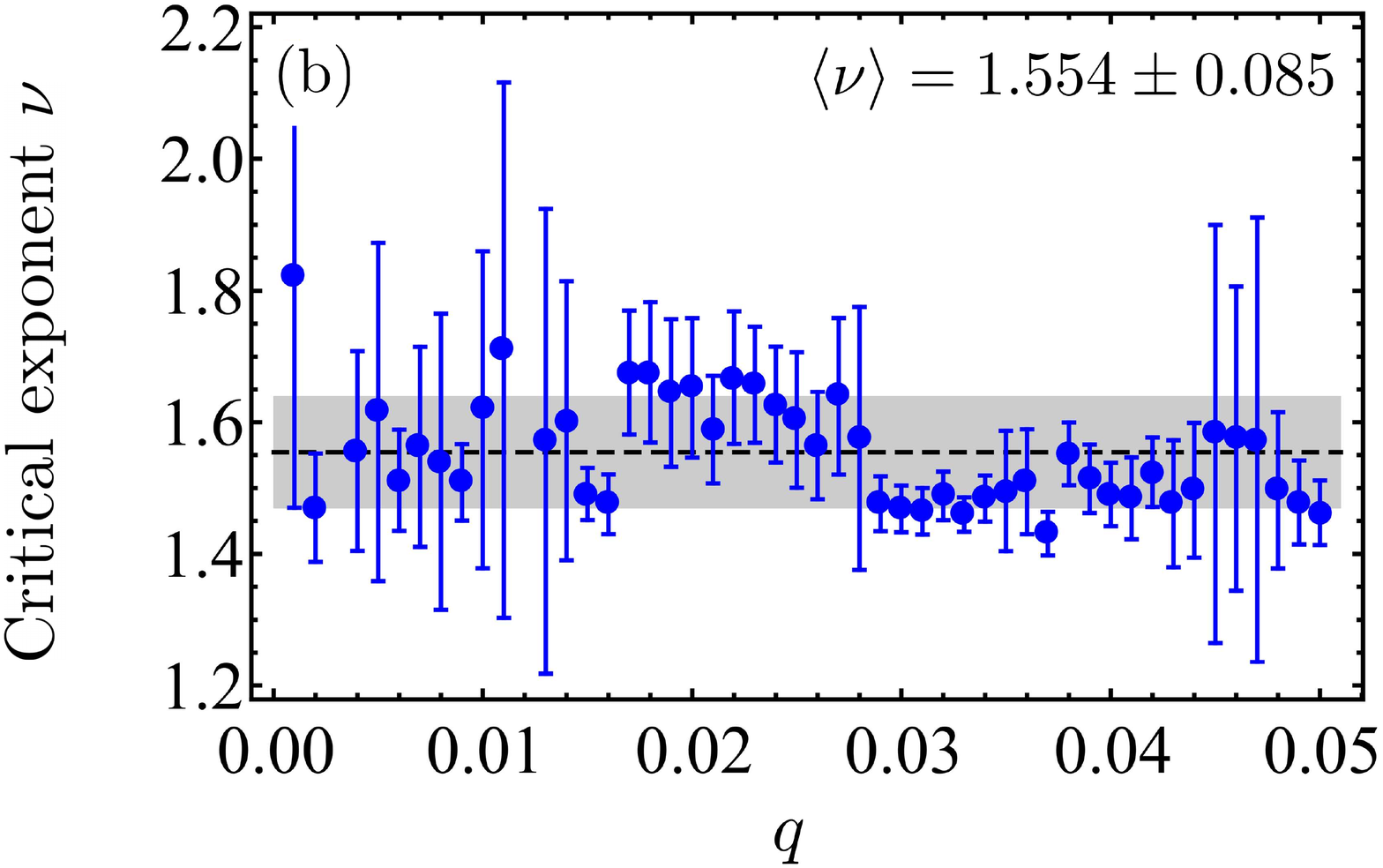}
\vspace{-5mm}
\caption{\label{fig_me_nu} Results for the model with resonant scattering. The mobility edge (a) and the critical exponent (b) determined from the fits to $q$-th percentiles, as functions of $q$. The horizontal dashed line shows the values of $\mathrm{Re} \Lambda_c$ and $\nu$ averaged over all $q = 0.001$--0.05. The grey areas correspond to the uncertainties of the averages.}
\end{figure}

In the following, we set $c_p/c_s = 2$ typical for metals (e.g., aluminum or tin) but also for some other materials, such as, e.g., polystyrene. {As discussed in Ref.\ \onlinecite{skip16},} the localization transition manifests itself in the statistical properties of the Thouless conductance $g = \mathrm{Im} \Lambda/\delta(\mathrm{Re}\Lambda)$, where $\delta(\mathrm{Re}\Lambda)$ is the average spacing between real parts of eigenvalues. We compute the probability density $p(\ln g)$ by averaging over many independent random configurations $\{ \vec{r}_m \}$.
{The number of different configurations that we use is adjusted to obtain at least $2 \times 10^7$ eigenvalues for $N \leq 10^4$ and $10^7$ eigenvalues for $N > 10^4$ \cite{note2}.} The small-$g$ part of $p(\ln g)$ tends to become independent of the size of the system at a mobility edge and can be used to determine the critical properties of the localization transition. To this end, we follow Ref.\ \onlinecite{skip16} and consider small-$q$ percentiles $\ln g_q$ of $p(\ln g)$.
{The percentiles $\ln g_q$ are defined by
\begin{eqnarray}
q &=& \int\limits_{-\infty}^{\ln g_q} p(\ln g) d(\ln g).
\label{perc}
\end{eqnarray}
}

When the density of scatterers is high enough, $\ln g_q$ become independent of the system size at critical points where localization transitions take place, as we illustrate in Fig.\ \ref{fig_percentiles}(a) for $q = 0.02$. We find two such points ($\mathrm{Re} \Lambda_c \simeq -1$ and $-2.4$ for $\rho/k_s^3 = 0.15$), with a band of localized states in between.
{Collective effects shift the latter with respect to the scattering resonance of individual scatterers corresponding to $\mathrm{Re} \Lambda = 0$. Although the exact position of this band cannot be predicted based on simple arguments, it is clear that disorder-induced localized states cannot exist far from the resonance frequency $\omega_0$, i.e. for $|\mathrm{Re} \Lambda| \gg 1$, because scattering by individual scatterers becomes negligible for $|\omega - \omega_0| \gg \Gamma_0$ and no collective effects can make a collection of such weak scatterers strongly scattering. By construction, our scattering medium becomes weakly scattering and then transparent as the frequency $\omega$ is detuned from $\omega_0$ by several $\Gamma_0$. It follows then that the number of mobility edges should be even in our model of identical resonant scatterers with a single narrow resonance.
Having two mobility edges near a resonance of individual scatterers} is typical for disordered samples used in recent experiments \cite{cobus16}, which confirms that our model is relevant for the interpretation of the latter.
{One should keep in mind, however, that spherical scatterers used in the experiments of Ref.\ \onlinecite{cobus16} have multiple resonances and that in this case, the link between resonant properties of individual scatterers and those of a large ensemble of them is not simple. For periodic arrangements of spherical scatterers, this problem has been discussed in Ref.\ \onlinecite{kafesaki95} and some arguments of that paper apply to disordered systems as well.}

The quality of our numerical data is considerably better around the first critical point {$\mathrm{Re} \Lambda_c \simeq -1$}, which we analyze in detail using the finite-size scaling approach. We follow the procedure described in Ref.\ \onlinecite{skip16} (see Ref.\ \onlinecite{slevin14} for a review of the finite-size scaling approach to Anderson localization). $\ln g_q$ is expanded in power series as a function of relevant and irrelevant scaling variables $\psi_1 = f_1(w)L^{1/\nu}$ (up to an order $n_1$) and $\psi_2 = f_2(w)L^{y}$ (up to an order $n_2$), where $f_j(w)$, in their turn, are expanded in series in $w = (\mathrm{Re} \Lambda - \mathrm{Re} \Lambda_c)/\mathrm{Re} \Lambda_c$ (up to orders $m_j$). The numerical data falling within $\pm \Delta$ of the estimated critical point $(\ln g_q)_c$ are fitted using $m_1 = n_1 \leq 4$; $m_2 = n_2 \leq 2$ with a requirement that the contribution of the irrelevant variable to the fit should not exceed 10\% of $\Delta$. For $\Delta = 2$, reasonable fits are obtained with $m_1 = n_1 = 3$ and $m_2 = n_2 = 1$, see Fig.\ \ref{fig_percentiles}(b).

Among all the fit parameters, the mobility edge $\mathrm{Re} \Lambda_c$ and the critical exponent $\nu$ are of main interest for us. Results obtained for very small $q$ suffer from large statistical errors because the limited number of random realizations does not allow to estimate $\ln g_q$ reliably. On the other hand, large-$q$ results are not trustworthy because the distributions $p(\ln g)$ following from our calculations start to violate the single-parameter scaling hypothesis when $\ln g$ is increased. Reliable results correspond to $q = 0.001$--0.05 shown in Fig.\ \ref{fig_me_nu}. Averaging over $q$ yields, in particular, an estimate of the critical exponent $\langle \nu \rangle  = 1.554 \pm 0.085$ which is in good agreement with the value $1.55 \pm 0.07$ following from the same analysis for scalar waves \cite{skip16}. In contrast, the mobility edge $\mathrm{Re} \Lambda_c$ turns out to be shifted to a higher frequency with respect to its value in the scalar model \cite{skip16}, similarly to the model with nonresonant scattering.

\subsection{Inverse participation ratio of quasi-modes}
\label{secipr}

\begin{figure}
\includegraphics[width=0.99\columnwidth]{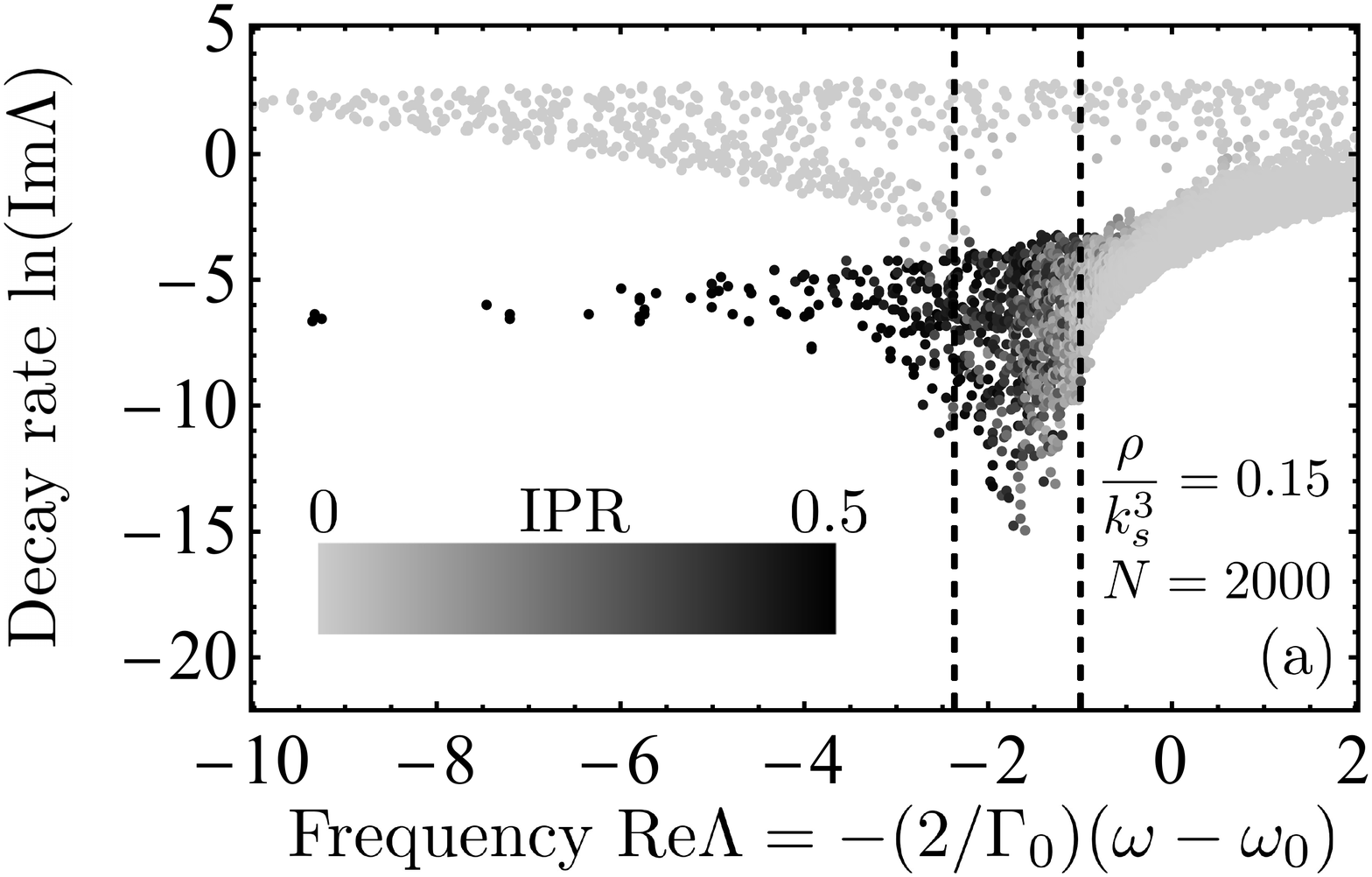}\\
\vspace{-7mm}
\includegraphics[width=0.99\columnwidth]{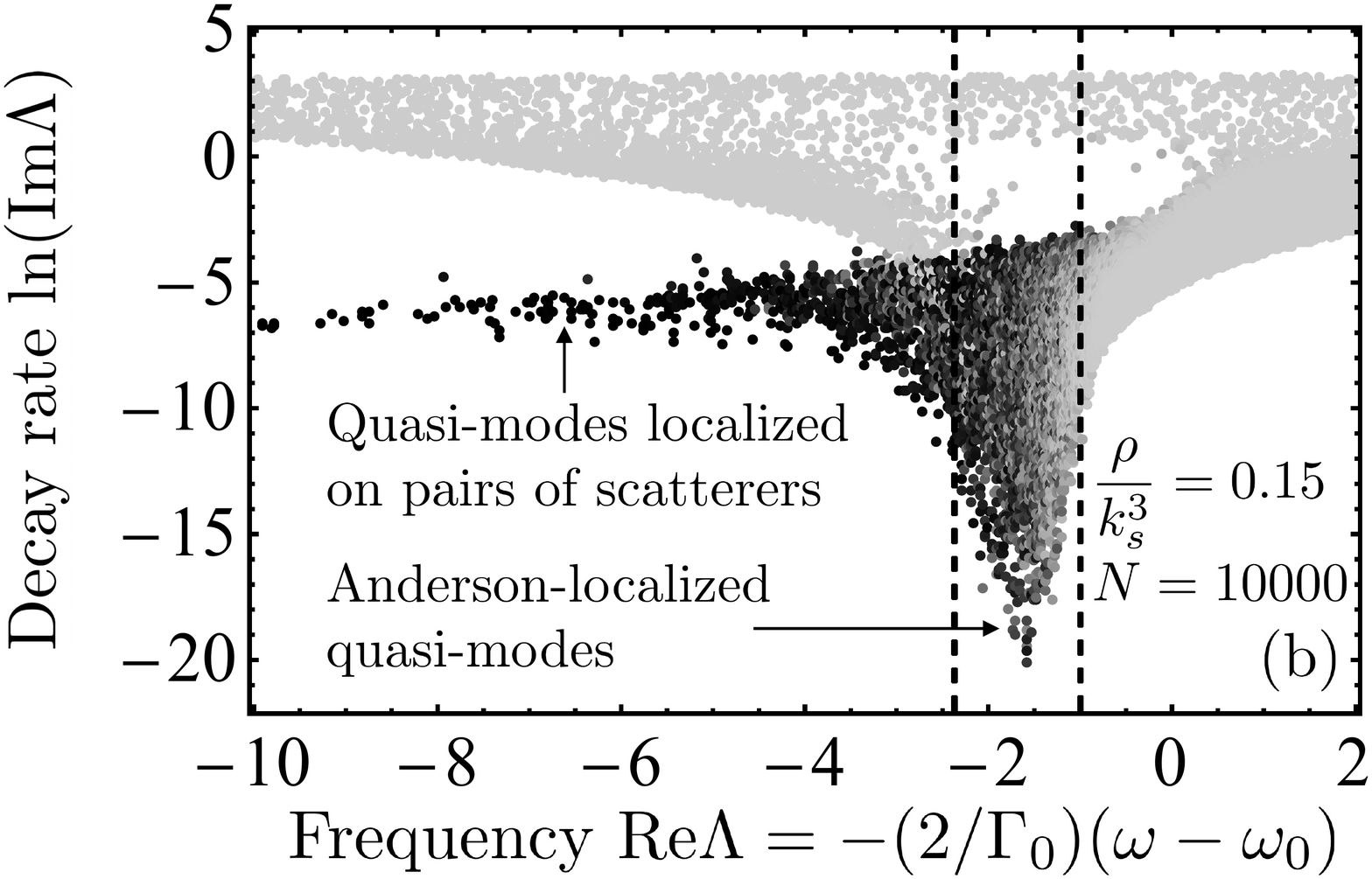}
\vspace{-5mm}
\caption{\label{fig_ipr}
{Eigenvalues of the Green's matrix (\ref{green}) are shown by dots for a single random configuration of $N$ scatterers  in a sphere at a fixed number density $\rho/k_0^3 = 0.15$. The grey scale of each dot corresponds to the IPR (\ref{ipr}) of the corresponding eigenvector. Fully extended eigenvectors have $\mathrm{IPR} \sim 1/N \ll 1$ and are shown in light grey. $N = 2000$ in the panel (a) and 10000 in the panel (b). Dashed vertical lines show the mobility edges.}}
\end{figure}

{As we explained already in Sec.\ \ref{secfss}, a \textit{quantitative} statistical analysis that would allow for estimating the mobility edges and the critical exponent of the localization transition with a sufficient accuracy based on the localization properties of quasi-modes, would be computationally too demanding. However, one can get a good \textit{qualitative} understanding of Anderson localization in the model described by the random matrix (\ref{green}) by considering its eigenvectors $| R_n \rangle$ obtained for a single random configuration $\{ \vec{r}_m \}$ of scatterers. The spatial localization of an eigenvector can be quantified by its inverse participation ratio (IPR):
\begin{eqnarray}
\mathrm{IPR}_n = \frac{\sum\limits_{m = 1}^N\left( \sum\limits_{\alpha = x, y, z} \left| R_n^{m, \alpha} \right|^2 \right)^2}{\left( \sum\limits_{m = 1}^N \sum\limits_{\alpha = x, y, z} \left| R_n^{m, \alpha} \right|^2 \right)^2},
\label{ipr}
\end{eqnarray}
where $R_n^{m, \alpha}$ denotes the component of the vector $| R_n \rangle$ on the scatterer $m$ with a polarization parallel to $\alpha = x$, $y$, or $z$. IPR varies from $1/N$ for a fully extended eigenvector to 1 for an eigenvector localized on a single scatterer. Typically, $\mathrm{IPR} \simeq 1/K$ for a state localized on $K$ scatterers. Figure \ref{fig_ipr} shows the eigenvalues of a random realization of the matrix (\ref{green}), with the grey scale of a dot showing an eigenvalue determined by the IPR of the corresponding eigenvector. Several comments are in order.
}

{In Fig.\ \ref{fig_ipr}, we clearly see the band of localized states between the mobility edges determined from the finite-size scaling of percentiles in Sec.\ \ref{secfss} [vertical dashed lines in Fig.\ \ref{fig_percentiles}(a)]. The spatial localization of quasi-modes corresponding to eigenvalues inside this band becomes stronger as the number of scatterers (and hence the system size $R \propto (N/\rho)^{1/3}$) increases, as one concludes by comparing the panels (a) and (b) of Fig.\ \ref{fig_ipr}, where results are presented for $N = 2000$ and $N = 10000$, respectively. At the same time, the decay rates of quasi-modes inside the band decrease exponentially with the system size. Such a scaling with system size is typical for Anderson localization and should be contrasted with the situation taking place for quasi-modes localized on pairs of closely located scatterers and corresponding to the eigenvalues belonging to the long ``tail'' with large negative $\mathrm{Re} \Lambda$ in Fig.\ \ref{fig_ipr}. Such quasi-modes have $\mathrm{IPR} \simeq 0.5$ and are hardly sensitive to the system size, as follows from the comparison of panels (a) and (b) of Fig.\ \ref{fig_ipr}. Their origin is the same as for scalar waves \cite{skip16} and light \cite{skip14}: waves emitted by a pair of closely located scatterers ($k_0 r_{mn} \ll 1$) oscillating out of phase interfere destructively and thus cancel each other. The corresponding quasi-mode effectively decouples from the ``outside world'', independently of the presence of other scatterers in the latter and their number $N$. Such ``subradiant'' localized quasi-modes exist at any scatterer density and number but do not show the scaling behavior associated with Anderson localization.
}

{
Figure \ref{fig_ipr} shows that the two mobility edges that we clearly see in Fig.\ \ref{fig_percentiles}(a) are significantly different in nature. Indeed, whereas the low-frequency mobility edge $\mathrm{Re} \Lambda \simeq -1$ corresponds to a sharp transition from extended quasi-modes with $\mathrm{IPR} \sim 1/N \ll 1$ to localized quasi-modes with $1/N \ll \mathrm{IPR} < 0.5$, localized modes are present on the both sides from the high-frequency mobility edge $\mathrm{Re} \Lambda \simeq -2.4$. However, as we discussed above, only the modes on the right from it are genuine Anderson-localized modes. Even though the scaling analysis of Fig.\ \ref{fig_percentiles}(a) makes a difference between the two different localization mechanisms and correctly detects a change in scaling behavior at the high-frequency mobility edge, the results presented in Fig.\ \ref{fig_percentiles}(a) turn out to be much more noisy near the high-frequency mobility edge than near the low-frequency one. As a result, a scaling analysis of the high-frequency localization transition does not yield conclusive results and we do not present it here.
}

\subsection{Comparison between elastic and electromagnetic waves}

The Green's matrix (\ref{green}) looks noticeably similar to its optical counterpart describing light scattering \cite{skip14}. In particular, it features near-field terms diverging as $1/r_{mn}^3$ at small $r_{mn}$. These terms associated with quasistatic dipole-dipole interactions were found to prevent Anderson localization of light \cite{skip14} but apparently do not play any important role in the elastic case. The mathematical reason for this is that these terms cancel out in Eq.\ (\ref{green})  for $k_{p,s} r_{mn} \ll 1$, leaving us with ${\hat G}_{mn} \propto 1/r_{mn}$.
{Indeed, the elastic Green's tensor (\ref{green0}) can be separated in contributions of compressive (longitudinal) and shear (transverse) waves that propagate independently of each other only in the far field $k_{p,s} \Delta r \gg 1$. In the near field $k_{p,s} \Delta r \ll 1$, the corresponding parts of the Green's tensor combine to give
\begin{eqnarray}
{\hat{\cal G}}(\mathbf{r}, \mathbf{r}',\omega) &=& \frac{1}{8 \pi \rho_0 c_s^2 \Delta r} \left\{
\left[1 + \left( \frac{c_s}{c_p} \right)^2 \right] \mathbb{1} \right.
\nonumber \\
&+& \left. \left[1 - \left( \frac{c_s}{c_p} \right)^2 \right] \frac{\Delta\mathbf{r} \otimes \Delta\mathbf{r}}{\Delta r^2}
\right\},
\label{greennf}
\end{eqnarray}
which diverges as $1/\Delta r$ only and hence is integrable in 3D.} This is very different from the optical case where
{${\hat{\cal G}} \propto 1/\Delta r^3$} and ${\hat G}_{mn} \propto 1/r_{mn}^3$ at small distances. The physical reason for elastic waves to behave differently from light stems from the existence of propagating longitudinal waves that get scattered and eventually localized in the same way as scalar waves do. In contrast, propagating waves are transverse in optics, whereas longitudinal fields give rise to dipole-dipole interactions between scatterers. Efficient only at small distances, these interactions open an additional, nonradiative channel of energy transport in sufficiently dense ensembles of scatterers \cite{nieuwen94}.

\section{Conclusions}
\label{sec:concl}

We studied the Anderson localization transition in two different models of elastic wave scattering in 3D: a model with nonresonant scattering, where the localization transition takes place upon increasing the frequency of a wave, and a model with resonant scattering, where a narrow band of localized states is separated from the rest of the spectrum by two mobility edges. We find that for both models, the vector character of waves shifts the localization transitions to higher frequencies with respect to the scalar case. At the same time, the critical exponent $\nu$ coincides with its value in the scalar case within the accuracy of our analysis, suggesting that the universality class of the transition does not change. Our best estimates of $\nu$ are $1.564 \pm 0.009$ and $1.554 \pm 0.085$ for the two models, respectively. They agree with the value $\nu \simeq 1.571$ in the 3D orthogonal universality class \cite{slevin14}. Hence, our results suggest that despite their vector character, elastic waves exhibit a disorder-induced localization transition of the same orthogonal universality class as the spinless electrons in a 3D disordered potential.

\begin{acknowledgements}
This work was funded by the Agence Nationale de la Recherche (project ANR-14-CE26-0032 LOVE). A part of the computations presented in this paper were performed using the Froggy platform of the CIMENT infrastructure ({\tt https://ciment.ujf-grenoble.fr}), which is supported by the Rhone-Alpes region (grant CPER07\verb!_!13 CIRA) and the Equip@Meso project (reference ANR-10-EQPX-29-01) of the programme Investissements d'Avenir supervised by the Agence Nationale de la Recherche. Y.M.B. acknowledges support of the Council of the President of the Russian Federation for Support of Young Scientists and Scientific Schools (project no. SP-3299.2016.1).
\end{acknowledgements}


\begin{thebibliography}{99}

\bibitem{anderson58}
P.W. Anderson,
Absence of Diffusion in Certain Random Lattices,
\textit{Phys. Rev.} \textbf{109}, 1492--1505 (1958).

\bibitem{abrahams10}
\textit{50 Years of Anderson Localization}, edited by E. Abrahams (Singapore, World Scientific, 2010).

\bibitem{abrahams79}
E. Abrahams, P.W. Anderson, D.C. Licciardello, and T.V. Ramakrishnan,
Scaling Theory of Localization: Absence of Quantum Diffusion in Two Dimensions,
Phys. Rev. Lett. \textbf{42}, 673--676 (1979).

\bibitem{evers08}
F. Evers and A.D. Mirlin,
Anderson Transitions,
Rev. Mod. Phys. \textbf{80}, 1355 (2008).

\bibitem{rosenbaum80}
T.F. Rosenbaum, K. Andres, G.A. Thomas, and R.N. Bhatt,
Sharp Metal-Insulator Transition in a Random Solid,
Phys. Rev. Lett. \textbf{45}, 1723 (1980).

\bibitem{paalanen82}
M.A. Paalanen, T.F. Rosenbaum, G.A. Thomas, and R.N. Bhatt,
Stress Tuning of the Metal-Insulator Transition at Millikelvin Temperatures,
Phys. Rev. Lett. \textbf{48}, 1284 (1982).

\bibitem{hu08}
H. Hu, A. Strybulevych, J.H. Page, S.E. Skipetrov, and B.A. van Tiggelen,
Localization of Ultrasound in a Three-Dimensional Elastic Network,
Nature Phys. \textbf{4}, 945 (2008).

\bibitem{cobus16}
L.A. Cobus, S.E. Skipetrov, A. Aubry, B.A. van Tiggelen, A. Derode, and J.H. Page,
Anderson Mobility Gap Probed by Dynamic Coherent Backscattering,
Phys. Rev. Lett. \textbf{116}, 193901 (2016).

\bibitem{chabe08}
J. Chab\'{e}, G. Lemari\'{e}, B. Gr\'{e}maud, D. Delande, P. Szriftgiser, and J.C. Garreau,
Experimental Observation of the Anderson Metal-Insulator Transition with Atomic Matter Waves,
Phys. Rev. Lett. \textbf{101}, 255702 (2008)

\bibitem{jendr12}
F. Jendrzejewski, A. Bernard, K. M\"{u}ller, P. Cheinet, V. Josse, M. Piraud, L. Pezz\'{e},	L. Sanchez-Palencia, A. Aspect and P. Bouyer,
Three-Dimensional Localization of Ultracold Atoms in an Optical Disordered Potential,
Nature Phys. \textbf{8}, 398 (2012).

\bibitem{vanderbeek12}
T. van der Beek, P. Barthelemy, P.M. Johnson, D.S. Wiersma, and A. Lagendijk,
Light Transport through Disordered Layers of Dense Gallium Arsenide Submicron Particles,
Phys. Rev. B \textbf{85}, 115401 (2012).

\bibitem{sperling16}
T. Sperling, L. Schertel, M. Ackermann, G.J. Aubry, C.M. Aegerter, and G. Maret,
Can 3D Light Localization be Reached in 'White Paint'?
New J. Phys. \textbf{18}, 013039 (2016).

\bibitem{skip16njp}
S.E. Skipetrov and J.H. Page,
Red Light for Anderson Localization,
New J. Phys. \textbf{18}, 021001 (2016).

\bibitem{skip14}
S.E. Skipetrov and I.M. Sokolov,
Absence of Anderson Localization of Light in a Random Ensemble of Point Scatterers,
Phys. Rev. Lett. \textbf{112}, 023905 (2014).

\bibitem{bellando14}
L. Bellando, A. Gero, E. Akkermans, and R. Kaiser,
Cooperative Effects and Disorder: A Scaling Analysis of the Spectrum of the Effective Atomic Hamiltonian,
Phys. Rev. A \textbf{90}, 063822 (2014).

\bibitem{escalante17}
J.M. Escalante and S.E. Skipetrov,
Longitudinal Optical Fields in Light Scattering from Dielectric Spheres and Anderson Localization of Light,
Ann. Phys. (Berlin) \textbf{529}, 1700039 (2017).

\bibitem{sheng94}
P. Sheng, M. Zhou, and Z.Q. Zhang,
Phonon Transport in Strong-Scattering Media,
Phys. Rev. Lett. \textbf{72}, 234 (1994).

\bibitem{chaudhuri10}
A. Chaudhuri, A. Kundu, D. Roy, A. Dhar, J.L. Lebowitz, and H.Spohn,
Heat Transport and Phonon Localization in Mass-Disordered Harmonic Crystals,
Phys. Rev. B \textbf{81}, 064301 (2010).

\bibitem{mohthus10}
C. Monthus and T. Garel,
Anderson Localization of Phonons in Dimension $d=1,2,3$: Finite-Size Properties of the Inverse Participation Ratios of Eigenstates,
Phys. Rev. B \textbf{81}, 224208 (2010).

\bibitem{pinski12}
S.D. Pinski, W. Schirmacher, T. Whall, and R.A. R\"{o}mer,
Localization-Delocalization Transition for Disordered Cubic Harmonic Lattices,
J. Phys.: Condens. Matter \textbf{24} 405401 (2012).

\bibitem{pinski12a}
S.D. Pinski, W. Schirmacher, and R.A. R\"{o}mer,
Anderson Universality in a Model of Disordered Phonons,
EPL \textbf{97} 16007 (2012).

\bibitem{amir13}
A. Amir, J.J. Krich, V. Vitelli, Y. Oreg, and Y. Imry,
Emergent Percolation Length and Localization in Random Elastic Networks,
Phys. Rev. X \textbf{3}, 021017 (2013).

\bibitem{beltukov17}
Y.M. Beltukov and S.E. Skipetrov,
Finite-Time Scaling at the Anderson Transition for Vibrations in Solids,
Phys. Rev. B \textbf{96}, 174209 (2017).

\bibitem{ludlam03}
J.J. Ludlam, S.N. Taraskin, and S.R. Elliott,
Disorder-Induced Vibrational Localization,
Phys. Rev. B \textbf{67}, 132203 (2003).

\bibitem{sepehrinia08}
R. Sepehrinia, M.R. Rahimi Tabar, and M. Sahimi,
Numerical Simulation of the Localization of Elastic Waves in Two- and Three-Dimensional Heterogeneous Media,
Phys. Rev. B \textbf{78}, 024207 (2008).

\bibitem{john83}
S. John, H. Sompolinsky, and M.J. Stephen,
Localization in a Disordered Elastic medium Near Two Dimensions,
Phys. Rev. B \textbf{27}, 5592 (1983).

\bibitem{photiadis17}
D.M. Photiadis,
Elastic Wave Transport in Disordered, Isotropic Media: A Supersymmetric Sigma Model,
Ann. Phys. (Berlin) \textbf{529}, 1600353 (2017).

\bibitem{huang09}
B.J. Huang and T.M. Wu,
Localization-Delocalization Transition in Hessian Matrices of Topologically Disordered Systems,
Phys. Rev. E \textbf{79}, 041105 (2009).

\bibitem{jacobs96}
D.J. Jacobs and M.F. Thorpe,
Generic Rigidity Percolation: The Pebble Game,
Phys. Rev. Lett. \textbf{75}, 4051 (1995).

\bibitem{skip16}
S.E. Skipetrov,
Finite-Size Scaling Analysis of Localization Transition for Scalar Waves in a Three-Dimensional Ensemble of Resonant Point Scatterers,
Phys. Rev. B \textbf{94}, 064202 (2016).

\bibitem{beltukov13}
Y.M. Beltukov, V.I. Kozub, D.A. Parshin,
Ioffe-Regel Criterion and Diffusion of Vibrations in Random Lattices,
Phys. Rev. B \textbf{87}, 134203 (2013).

\bibitem{taraskin99}
S.N. Taraskin and S.R. Elliott,
Low-Frequency Vibrational Excitations in Vitreous Silica: the Ioffe–Regel Limit,
J. Phys.: Condens. Matter \textbf{11}, A219 (1999).

\bibitem{landau}
L.D. Landau and E.M. Lifshitz,
\textit{Theory of Elasticity} (Pergamon Press, Oxford, 1970).


\bibitem{ben81}
A. Ben-Menahem and S.J. Singh,
\textit{Seismic Waves and Sources} (Springer-Verlag, New York, 1981).

\bibitem{snieder02}
R. Snieder,
General Theory of Elastic Wave Scattering,
in \textit{Scattering and Inverse Scattering in Pure and Applied Science},
edited by R. Pike and P. Sabatier (Academic Press, San Diego, 2002), pp. 528--542.

\bibitem{foldy45}
L.L. Foldy,
The Multiple Scattering of Waves: I. General Theory of Isotropic Scattering by Randomly Distributed Scatterers,
Phys. Rev. \textbf{67}, 107 (1945).

\bibitem{lax51}
M. Lax,
Multiple Scattering of Waves,
Rev. Mod. Phys. \textbf{23}, 287 (1951).

\bibitem{skip15}
S.E. Skipetrov and I.M. Sokolov,
Magnetic-Field-Driven Localization of Light in a Cold-Atom Gas,
Phys. Rev. Lett. \textbf{114}, 053902 (2015).

\bibitem{skip18}
S.E. Skipetrov,
Localization Transition for Light Scattering by Atoms in an External Magnetic Field,
arXiv:1805.04314.

\bibitem{kafesaki95}
M. Kafesaki and E.N. Economou,
Interpretation of the Band-Structure Results for Elastic and Acoustic Waves by Analogy with the LCAO Approach,
Phys. Rev. B \textbf{52}, 13317 (1995).

\bibitem{slevin14}
K. Slevin and T. Ohtsuki,
Critical Exponent for the Anderson Transition in the Three-Dimensional Orthogonal Universality Class,
New J. Phys. \textbf{16}, 015012 (2014).

\bibitem{nieuwen94}
Th.M. Nieuwenhuizen, A.L. Burin, Yu. Kagan, and G.V. Shlyapnikov,
Light Propagation in a Solid with Resonant Atoms at Random Positions,
Phys. Lett. A \textbf{184}, 360--365 (1994).

\bibitem{note1}
We attempted fits using $m_1$, $n_1$ up to 3 and $m_2$, $n_2$ up to 2 but found that increasing $m_1$, $n_1$ or introducing the irrelevant variable by setting $m_2$, $n_2 > 0$  improves the quality of fits only marginally while increasing the uncertainties of the values of best-fit parameters.

\bibitem{note2}
{The actual number of independent random configurations $\{ \vec{r}_m \}$ that we use is
3400 (for $N = 2000$),
1716 ($N=4000$),
1126 ($N=6000$),
862 ($N=8000$),
689 ($N=10000$),
300 ($N=12000$),
255 ($N=14000$), and
209 ($N=16000$).}

\end{thebibliography}
\end{document}